\documentclass[10pt,conference]{IEEEtran}
\usepackage[pdftex]{graphicx}
\usepackage{balance}
\ifCLASSINFOpdf
\usepackage{subfigure}
\else
\fi

\usepackage[cmex10]{amsmath}

\renewcommand\IEEEkeywordsname{Index Terms}
\begin{document}
	\title{Study of the US Road Network based on Social Network Analysis}
	
	\author{\IEEEauthorblockN{Elie Ngomseu Mambou$^{\ast}$, Samuel Nlend$^{\ast}$ and Harold Liu$^{\dagger}$}
		\IEEEauthorblockA{University of Johannesburg, Auckland Park, 2006, South Africa\\
			Beijing Institute of Technology, Beijing 100081, PR China\\
			Emails: emambou@uj.ac.za$^{\ast}$, samueln@uj.ac.za$^{\ast}$,  chiliu@bit.edu.cn$^{\dagger}$}} 
	
	\IEEEoverridecommandlockouts
	\IEEEpubid{\makebox[\columnwidth]{978-1-5386-2775-4/17/\$31.00~
	\copyright2017
	IEEE \hfill} \hspace{\columnsep}\makebox[\columnwidth]{ }}
	\maketitle

	\begin{abstract}
The complexity of big data structures and networks demands more research in terms of analysing and representing data for a better comprehension and usage. In this regard, there are several types of model to represent a structure. The aim of this article is to use a social network topology to analyse road network for the following States in the United States (US): California, Pennsylvania and Texas. Our approach mainly focuses on clustering of road network data in order to create ``communities''.    
	\end{abstract}
	
	\begin{keywords}
		
Social network analysis (SNA), Community structure, Spark, Road network, Tableau, MLib, GraphX.
		\end{keywords}

	\IEEEpeerreviewmaketitle

	\section{Introduction}
The social network analysis mainly referred to as SNA is the study or process of analysing big data structures or network via graphical representation through techniques such as theory of graphs and network virtualization. This consists of classifying a network structure in \textit{nodes}, which are actual positions or entity within the network and \textit{edges} that are connections, relationships or links between nodes.   	
SNA are used in many fields such as economics, computer science, data mining, network modelling and sampling, biomedicine, and artificial intelligence.

In \cite{c1}, an investigation of nodes for large real-world social and information networks is presented. This study claims that large communities or clusters tend to be blended in more the rest of the network. Therefore according to interaction graphs within large networks, best communities in terms of conductance (how quick the network becomes an uniform distribution) are not likely to remain as the network grows in size.    

In \cite{c2}, a SNA techniques is presented for road network analysis. Three types of centralities are evaluated: degree centrality (used to detect central nodes in terms of spreading information and influencing the immediate neighbourhood), closeness centrality (average length of all shortest paths from a node to all other nodes, it appreciates the time or speed to reach other nodes from a starting position) and betweenness centrality (number of shortest paths passing through a node divided by all shortest paths in a network, it helps detecting places where the network would break apart). This property analysis helped to compare distribution in node centrality between down-town and residential areas within the network. The degree and entropy between the two areas are almost similar due to the planar network; while the distribution in closeness and betweenness are quite different due to network topology between the two areas.

Various techniques to represent SNA were reviewed in \cite{c3}, through several aspects of the network structure. It stipulates that the Web can be viewed as a social networks where documents are nodes of the sociogram (graphical representation of SNA) and edges are connections between documents. Weblogs as subset of Web can also be considered as social networks.   

A generalization of \cite{c1} is presented in \cite{c4}, where statistical properties for community-like sets of nodes are discussed for large real-world social and information networks. The main claim is that for a size scale of 100 nodes, there is a good network structure within the SNA; however beyond that range, the conductance within the network gets worse and the amount of community-like decreases.

A new algorithm to optimize the modularity of a network topology within a community structure was discussed in \cite{c5}. This algorithm performs at $\mathcal{O}(md \log n)$ where $n$ represents the number of vertices, $m$ the number of edges within the graph and $d$ the depth of the graph. This algorithm is applicable to large network and allows the extension of the community structure in SNA.

GIS (geographical information system) techniques were used in \cite{c6} to analyse the road network of Chandigarh city in India, for defining the optimal area for various facilities like schools, hospitals and fire stations. This was achieved through Google earth and then geo-referenced those facility areas. The network analysis for efficiency facility location was evaluated in terms of distance and time. This was helpful to appreciate the repartition or spatial allocation of various service areas within the city.  

In \cite{c7}, some methods were proposed to evaluate the accessibility of urban road network with a case study on Foshan city. The analysis of the road network was done through GIS spatial technology and some road network accessibility evaluation models were proposed and discussed from this analysis. The model was based on three evaluation indicators: the shortest time distance (STD), which is the minimum time spending route between one node and all other nodes within the road network, a lower STD indicates higher node's accessibility; the weighted average travel time (WATT), that is the sum of times from one node to all other nodes through the shortest time spending route, the WATT indicates how optimal a node's position is within the road network; and the accessibility index, this is a normalised index for the STD and the WATT, to estimate the actual accessibility of each node.

In \cite{c8}, a study was carried out to analyse the road network connectivity and spatial pattern in Calicut city in India based on spatial techniques. It was observed that the transport network fractality is very related with connectivity and coverage; the network density is the best indicator for predicting the fractality of a road network, therefore the level of road network development and the network spatial structure are extremely dependant. 

A road network analysis using geoinformatic techniques was proposed in \cite{c9} for Akola city. A system was provided based on GIS technology to mitigate disasters such as flood, fire and car accident within the study area; in a sense that the road network can be used to address each disaster within a short period of time through accessibility to facilities such as police station, fire station, ambulance and hospital.

In this article, a study of the US road is discussed based on SNA. We have used the data of the road network for three big States in the US which are California, Pennsylvania and Texas. The rest of this paper is structured as follow: In section II, the project overview is discussed with the tools to be used. Section III deals with the results and analysis of the road networks and finally Section IV summarises our achievements and presents some future directions.

\section{Project Overview}
Table \ref{tab1} presents the road network data to be analysed. This was retrieved from the Stanford large network dataset collection webpage, https://snap.stanford.edu/data/\#road. Nodes represent endpoints or intersections while edges are roads or links between nodes.

\begin{figure}[h]
\centering
\includegraphics[width=0.8\linewidth]{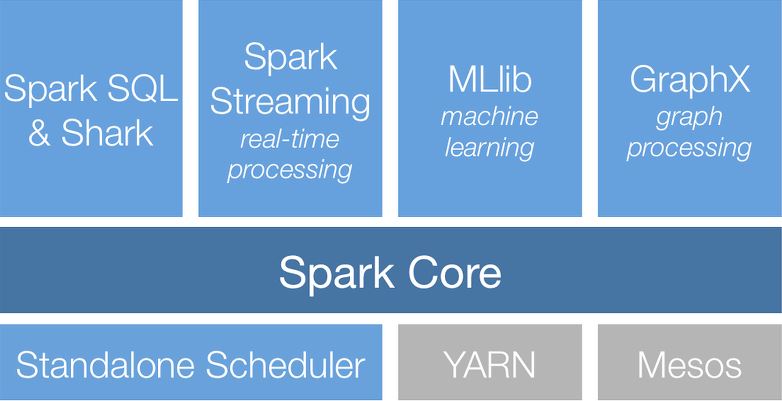}
\caption[Spark Technology]{Spark Technology}
\label{fig:spark}
\end{figure}

\begin{table}[h]
	\centering
	\caption{Road network data}\label{tab1}\vspace{-8pt}
\begin{tabular}{|c|c|c|c|c|}
	\hline 
	Name & Type & Nodes & Edges & Description \\ 
	\hline 
	roadNet-CA & Undirected & 1,965,206 & 2,766,607 & Road network of California \\ 
	\hline 
	roadNet-PA & Undirected & 1,088,092 & 1,541,898 & Road network of Pennsylvania \\ 
	\hline 
	roadNet-TX & Undirected & 1,379,917 & 1,921,660 & Road network of Texas \\ 
	\hline 
\end{tabular}
\end{table} 
Our analysis are based on the Spark Technology as shown in Fig. \ref{fig:spark}. This is computational framework for general purpose. The Spark streaming receives streams from large input sources then process them into clusters then push the results into databases or dashboards. It uses the \textit{MapReduce} (programming model to generate clusters through processing of big data) for scalability, fault tolerance, data locality based computations. The main advantages of this technology is that it performs parallel computation for big data analysis and works on distributed memory for better performance and iterative algorithm.

The system architecture for the SNA of these road networks was as follow: 
\begin{itemize}
	\item Operating system: Ubuntu 14.04.3 LTS
	\item Engine: Spark 1.5.2 using Scala 2.10.4
	\item Spark libraries: Spark streaming, Tableau, GraphX and MLib
	\item File storage: HDFS (Hadoop distributed file system).
	\item Other tools: sbt (open source framework for Scala projects) that allows interactive build tool.
\end{itemize} 

\section{Results and Analysis}
\begin{figure}[h]
\centering
\subfigure[]{\includegraphics[width=0.62\linewidth]{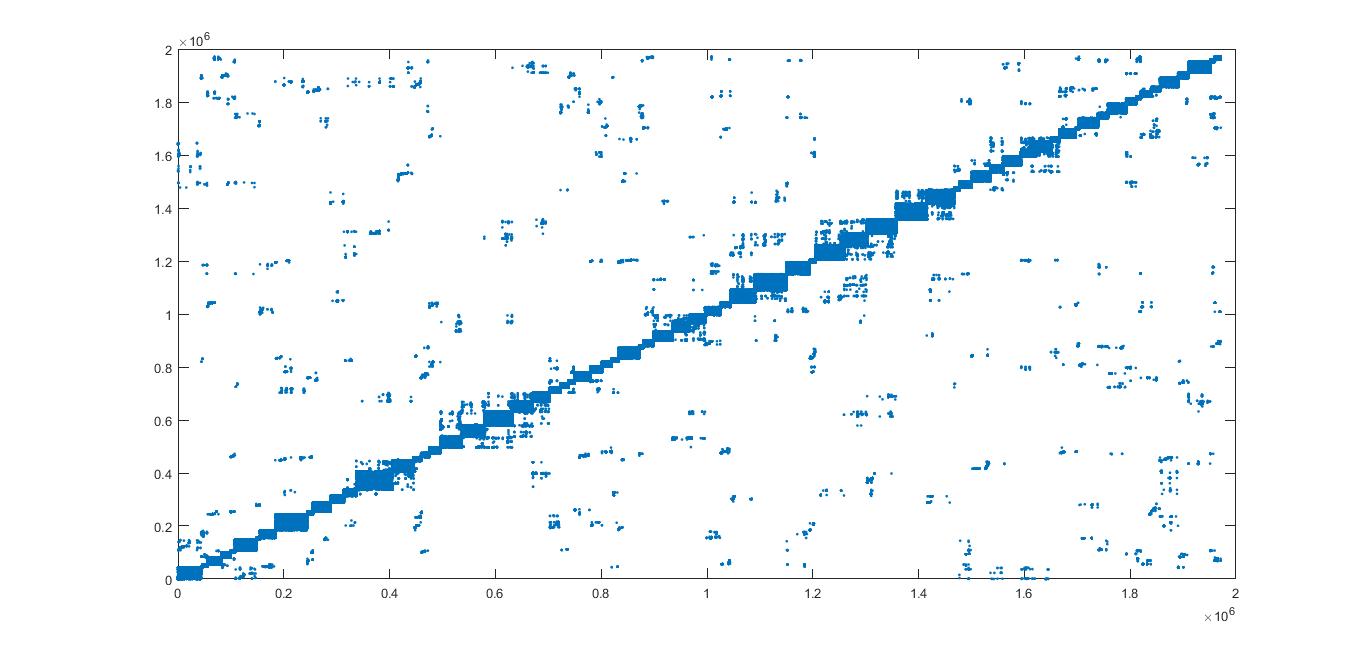}}
\subfigure[]{\includegraphics[width=0.62\linewidth]{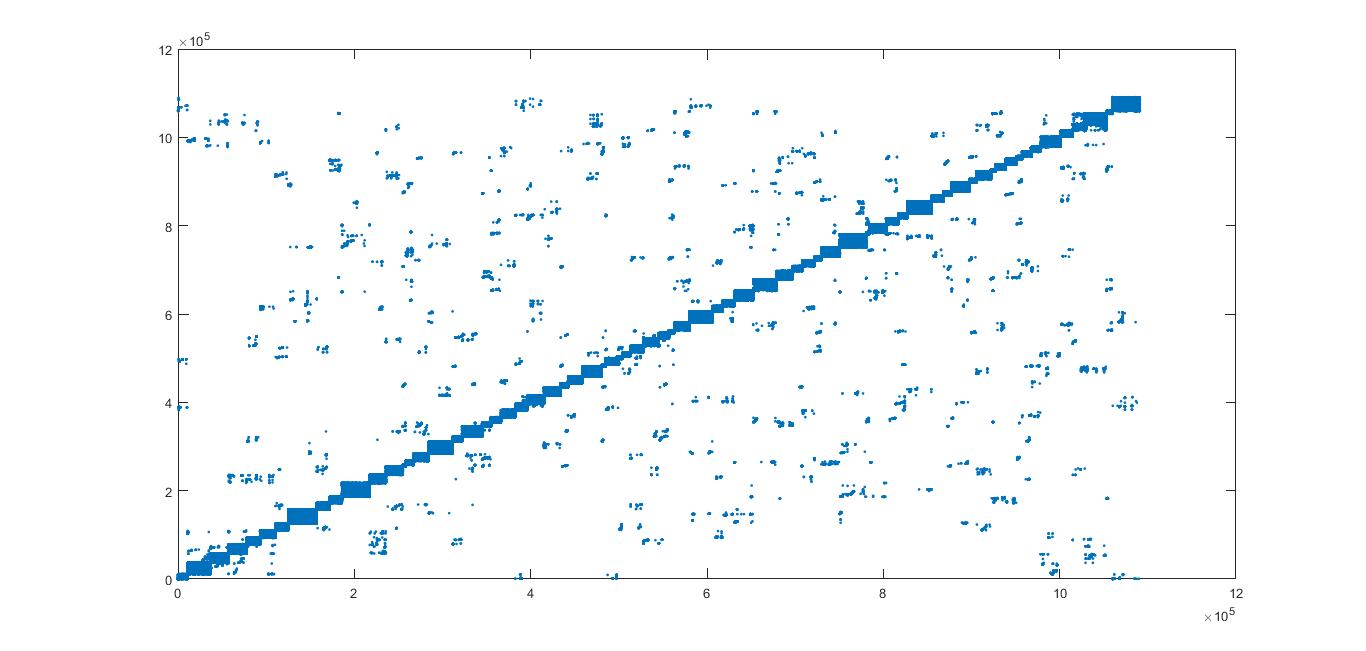}}
\subfigure[]{\includegraphics[width=0.62\linewidth]{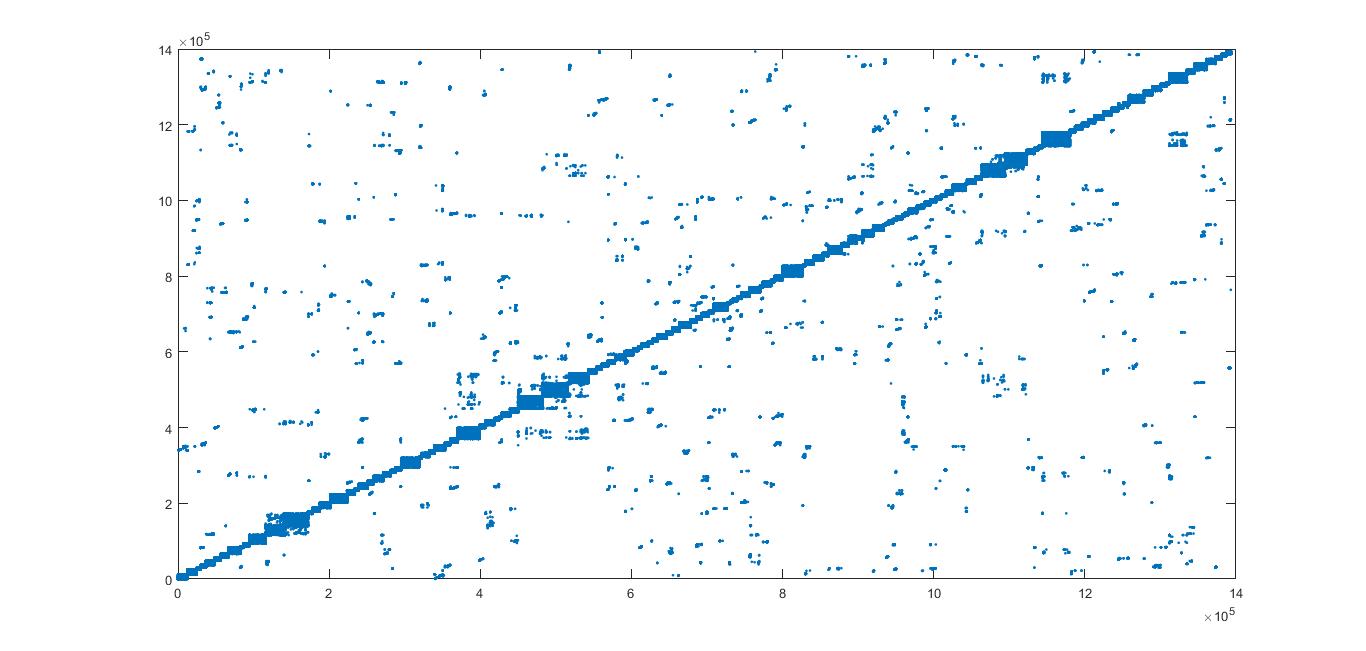}}
\caption{Road network nodes for (a)-California, (b)-Pennsylvania and (c)-Texas}
\label{fig:cpt1}
\end{figure}

\begin{figure}[h]
	\centering
	\subfigure[]{\includegraphics[width=0.62\linewidth]{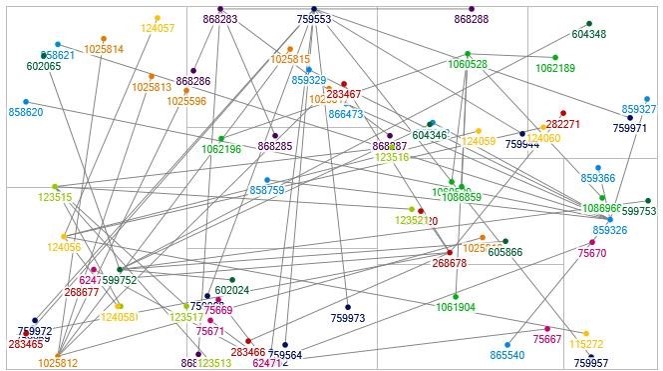}}
	\subfigure[]{\includegraphics[width=0.62\linewidth]{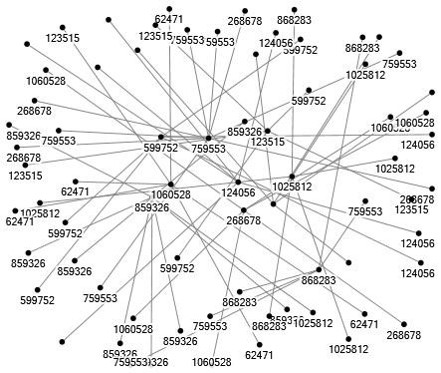}}
	\caption{Road network edges for (a)-Pennsylvania and (b)-Texas}
	\label{fig:cpt2}
\end{figure}

 If we call the graphical representation of a road network as $P$, then $P=(N,E)$ where $N$ is the set of nodes and $E$, the set of egdes within the graph $P$. Fig. \ref{fig:cpt1} presents the nodes of the network road for California, Pennsylvania and Texas. This is a simple scatter plot of $N$ vs. $E$, to visualize the topography of the road networks. The first observation on Fig. \ref{fig:cpt1} is all the three states present a similar road network topology in the sense that there is a strong concentration of nodes through the diagonal of the structure, this could represent the urban areas with main axes within a state and the surroundings are less developed areas or peripheries. Fig~\ref{fig:cpt2} presents the road network intersections and edges for Pennsylvania and Texas. This shows the actual links between nodes.

\subsection{GraphX and Spark Streaming}
In the theory of graph, the degree represents the number of edges incident to a node. In the case of the provided road network of type undirected, the edges are unidirectional. Given a node within $P$, the number of edges heading towards a node is called \textit{indegree} and the number of edges leaving a node is referred to as \textit{outdegree}. This analysis allows us to visualize how degree can affect the topology of a network road.

\begin{figure}[h]
\centering
\includegraphics[width=0.7\linewidth]{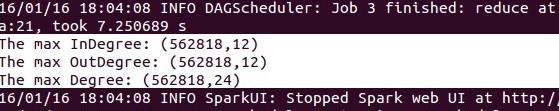}
\caption{Degree analysis for California road network}
\label{fig:p2}
\end{figure}

\begin{figure}[h]
\centering
\includegraphics[width=0.7\linewidth]{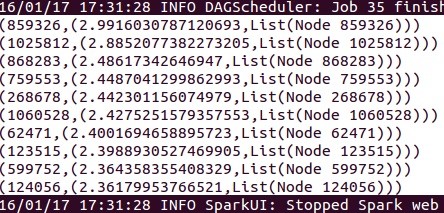}
\caption{Top 10 nodes with most page rank}
\label{fig:pr3}
\end{figure}

\begin{figure}[h]
\centering
\includegraphics[width=0.7\linewidth]{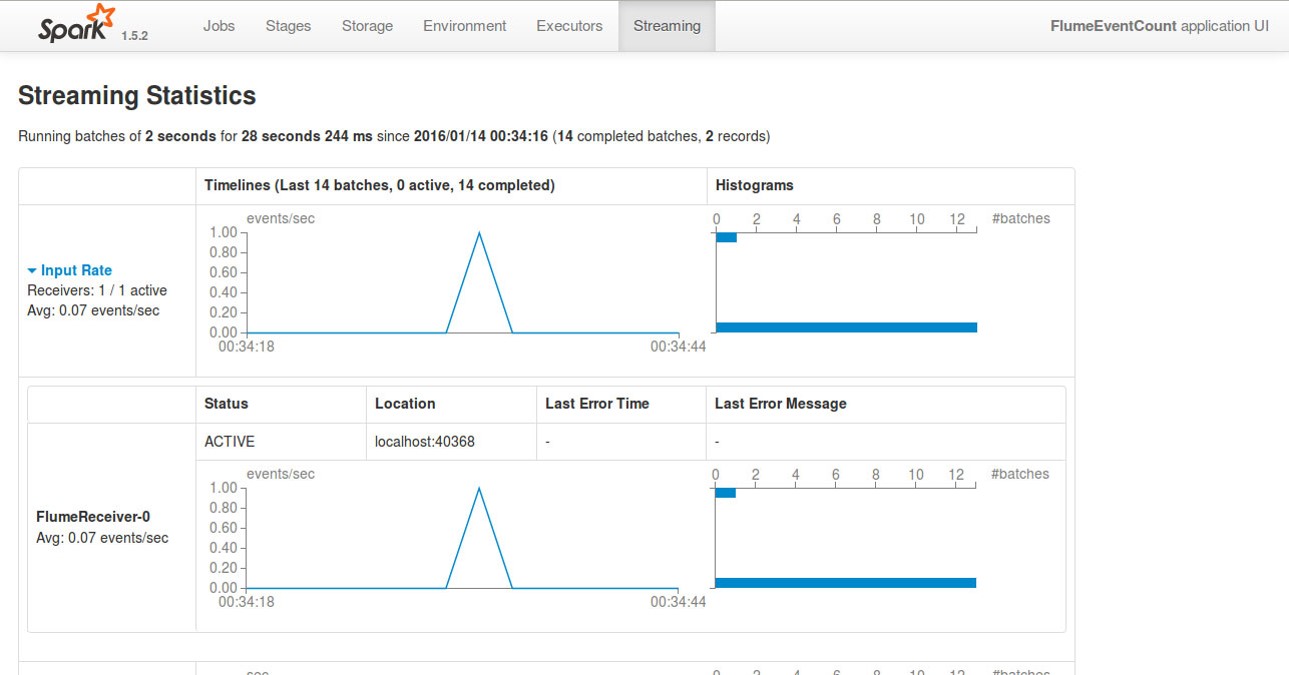}
\caption{streaming for top page rank intersections}
\label{fig:st}
\end{figure}

Fig. \ref{fig:p2} presents the node with the maximum degree, indegree and outdegree for the California road network. 
The page rank (PR) is a Google tool which counts the number and quality of edges to a node in order to estimate the importance of that node.  Fig. \ref{fig:pr3} shows the top 10 nodes with the highest page rank in the Pennsylvania road network. The format of the list is as follow: (Node Identity, Page Rank, List(Node attribute)). This shows the 10 intersections with the highest edges. Fig. \ref{fig:st} presents the live streaming statistics for the top page rank nodes.

\subsection{Data visualization via Tableau Software}
Tableau is a smart tool that allows data visualization of big networks.

\begin{figure}[h]
	\centering
	\subfigure[]{\includegraphics[width=0.7\linewidth]{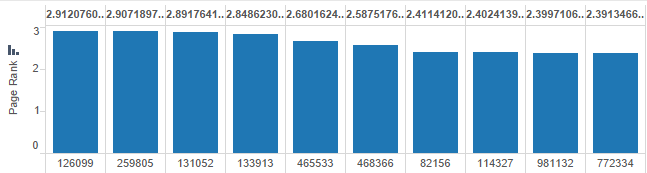}}
	\subfigure[]{\includegraphics[width=0.7\linewidth]{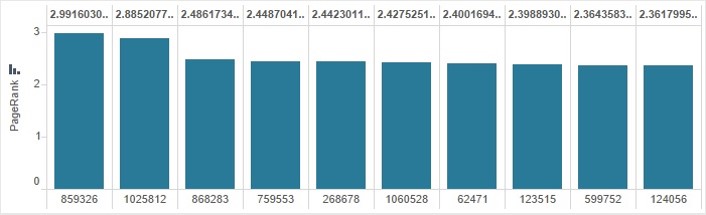}}
	\caption{Top 10 road endpoints in (a)-Pennsylvania vs. (b)-Texas}
	\label{fig:tb}
\end{figure}

\begin{figure}[h]
	\centering
	\subfigure[]{\includegraphics[width=0.6\linewidth]{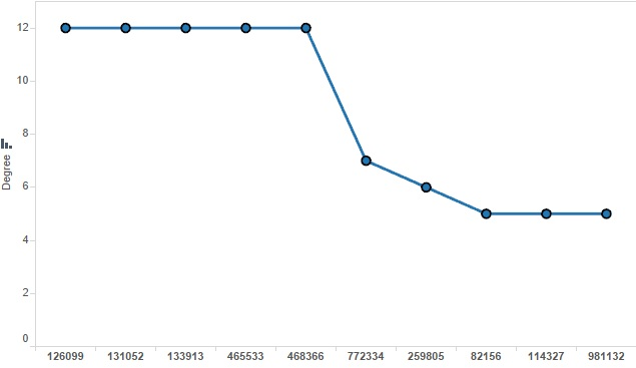}}
	\subfigure[]{\includegraphics[width=0.6\linewidth]{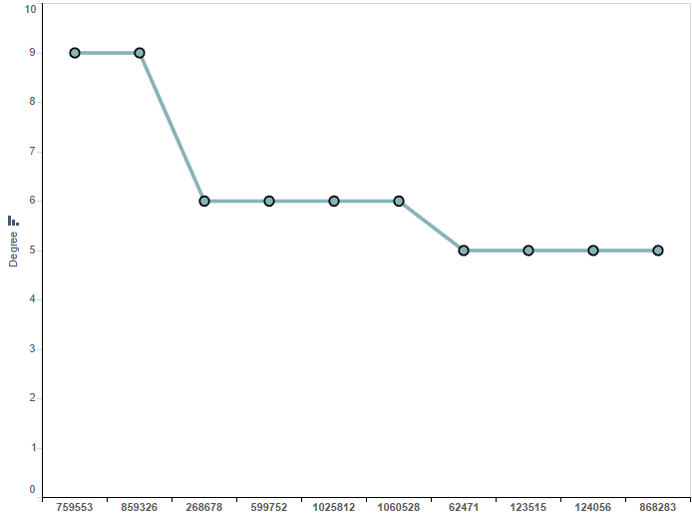}}
	\caption{Top 10 road endpoints in degree for (a)-Pennsylvania vs. (b)-Texas}
	\label{fig:tba}
\end{figure}

Fig.~\ref{fig:tb} shows the diagram of the top 10 road endpoints in Pennsylvania vs. Texas. We observed that the top 10 road nodes in Pennsylvania have greater page rank than that of Texas. Fig. \ref{fig:tba} is the representation of the top 10 road intersections in terms of degree for Pennsylvania versus Texas as in Fig.~\ref{fig:tb}. According to these two previous graphs, we observed that degree increases proportionally with page rank this is why the top 10 road intersection degree for Pennsylvania are higher than that of Texas.

\subsection{Clustering using MLlib}
MLlib is an Apache Spark's scalable machine learning library, it contains sub-libraries like Spark.mllib, Spark.ml(v1.2) compatible with many languages like Python, Java and Scala. It provides functionalities such as classification, regression, clustering, collaborative filtering  and dimensionality reduction. 

\begin{table}[h]
\centering
\includegraphics[width=0.7\linewidth]{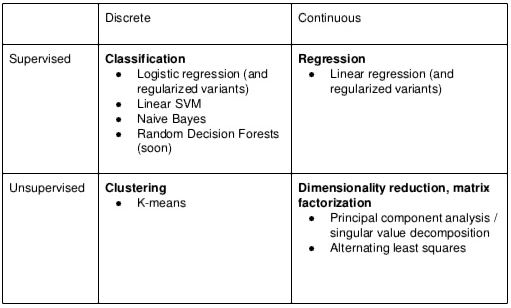}
\caption{MLlib tools}
\label{tbLib}
\end{table}
Table~\ref{tbLib} presents some tools for MLlib classified in terms of application properties: supervised/unsupervised and Discrete/continuous.

We will use the MLlib tool to essentially perform the \textit{k-means}  clustering on the road network, which is a very used asset in machine learning. The goal of the k-means clustering is to partition the graph $P=(N,E)$ into $k$ communities or clusters. Therefore each node belongs to the cluster that has its centre (centroid) to the nearest mean from that node. There are several algorithms to perform k-means.   

Let $P=(N,E)$ be the graphical representation of a road network. Each observation in the graph can be viewed as a point $p$ with two dimensional coordinates from $N$ and $E$, then  $P=\{p_1,p_2,\dots, p_t\}$ where $t$ is the number of elements in $N$. The k-means tool partition the $t$ points into $k$ communities where $k\leq t$.

The purpose of this algorithm is to minimize  the objective function:
\begin{equation}
G(\mathcal{A}_c)=\sum_{i}^{A}\sum_{j}^{A_i}(\parallel p_i-a_i\parallel)^2
\end{equation} 
Where $\parallel p_i-a_i\parallel$ is the euclidean distance between $p_i$, a point in $P$ and $a_i$ which is node in the set of centroids. $A_i$ is the number of points in cluster $i$ and $A$ is the number of centroids, $(A=k)$.

The set of centroids is such that $\mathcal{A}_c=\{a_1, a_2, \dots,a_A\}$. The algorithm consists of selecting A centroids and then assigns each point in $P$ to the closest centroid. After each assignment, the subsequent centroid is updated as follow:
\begin{equation}
a_i=\frac{1}{A_i}\sum_{z=1}^{A_i}p_z
\end{equation}
Finally the distance between point against the new centroids is recalculated and the process repeats until all data points are assigned to a cluster.

This process is robust and quick, it performs at $\mathcal{O}(2kts)$ with $k$ being the number of clusters, $t$, the number of points in $P$, $s$ is the number of iterations  and $2$ is the dimension for every point.

\begin{figure}[h]
\centering
\includegraphics[width=0.7\linewidth]{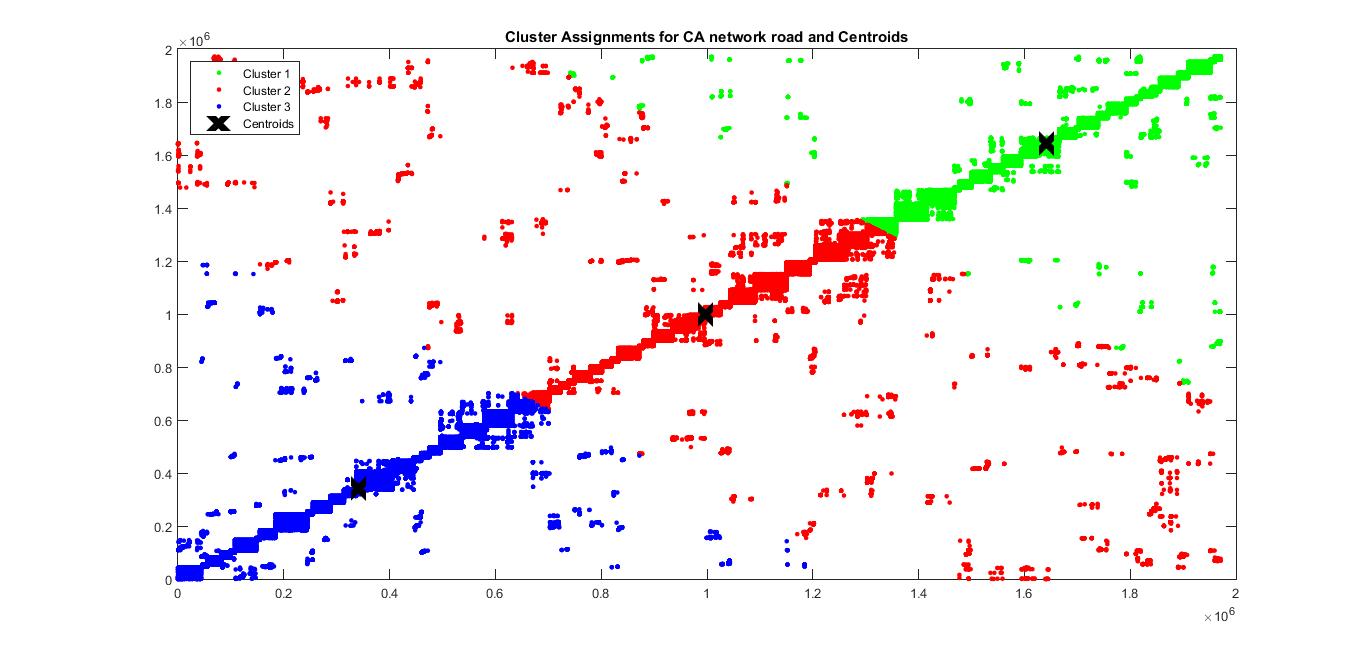}
\caption{$k$-means clustering on California road network with $k=3$}
\label{fig:cacl1}
\end{figure}

\begin{figure}[h]
\centering
\includegraphics[width=0.7\linewidth]{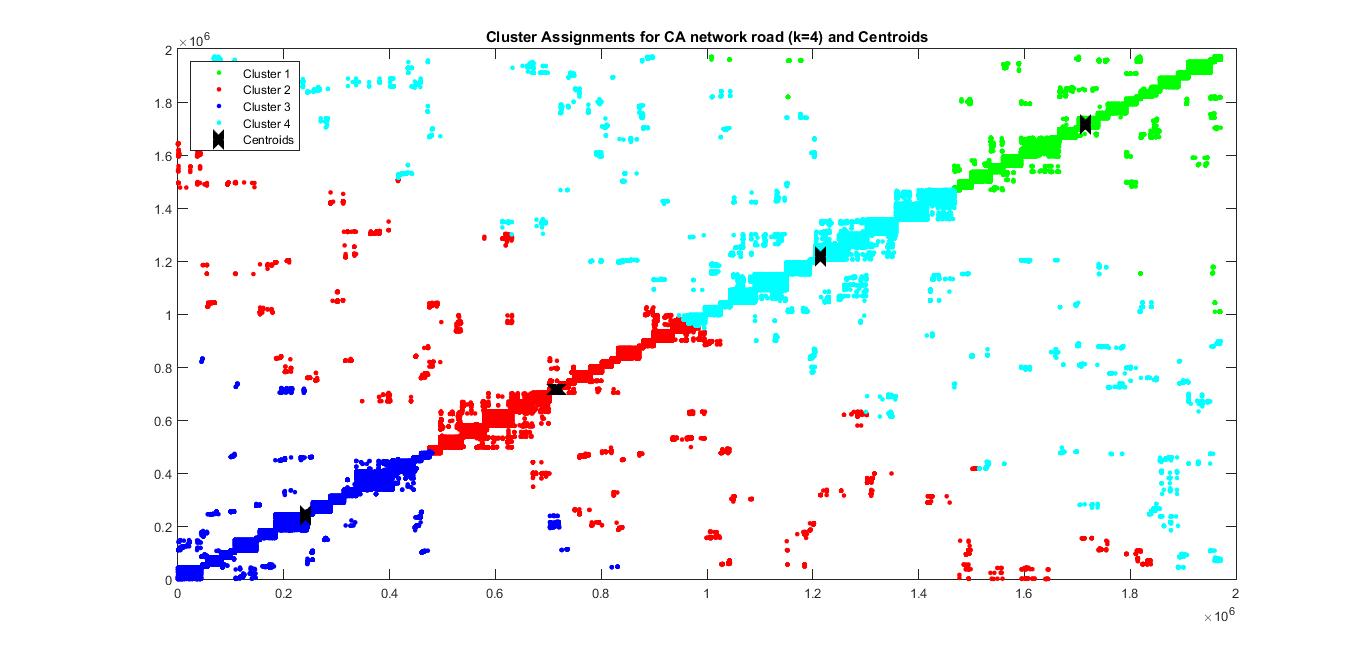}
\caption{$k$-means clustering on California road network with $k=4$}
\label{fig:cacl2}
\end{figure}

Figs. \ref{fig:cacl1} and \ref{fig:cacl2} show the $k$-means clustering for the California road network with $k=3$ and $k=4$ respectively. The cross on those graphs represents the position of the centroid. 

However the $k$-means analysis presents less flexibility because $k$ should be known beforehand, also this algorithm can not make use of non-integer data points or random euclidean distances, therefore it is not suitable for prediction and probabilistic based applications. There are some improvements that have been proposed for the $k$-mean such as Fuzzy means clustering, $k$-medoids, $k$-means++ or $k$-medians clustering that give a better analysis and performance.

\section{Conclusion and future work}
In this article, we achieved some data analysis trough GraphX, basic real time display with Spark streaming, network road virtualization using Tableau and clustering performed via MLlib libraries. Spark framework allowed a deep analysis of big data such as the US Road network. This analysis can be extended to data classification, estimation and
quantization.

\section{Acknowledgment}
The authors would like to thank the group of advanced software engineering course students 2015/2016 at the software school of the Beijing Institute of Technology for their contribution.

\balance

\end{document}